\documentclass[preprint,aps,draft]{revtex4}
\usepackage{graphicx}
\usepackage{dcolumn}
\usepackage{bm}

\begin{document}

\title{Microwave measurements of the photonic bandgap\\
in a two-dimensional photonic crystal slab}
\author{J. M. Hickmann}
\altaffiliation[Also at ]{Departamento de F\'{\i}sica, Universidade Federal de Alagoas, Cidade
Universit\'{a}ria, 57072-970, Macei\'{o}, AL, Brazil.}
\author{D. Solli}
\author{C. F. McCormick}
\author{R. Plambeck}
\altaffiliation[Also at ]{Astronomy Department; University of California; Berkeley, CA 94720-3411.}
\author{R. Y. Chiao}
\affiliation{Department of Physics; University of California; Berkeley, CA 94720-7300.}

\begin{abstract}
We have measured the photonic bandgap in the transmission of microwaves
through a two-dimensional photonic crystal slab. The structure was
constructed by cementing acrylic rods in a hexagonal closed-packed array to
form rectangular stacks. We find a bandgap centered at approximately 11 GHz,
whose depth, width and center frequency vary with the number of layers in
the slab, angle of incidence and microwave polarization.
\end{abstract}

\maketitle



There has been much recent interest in photonic bandgap materials \cite%
{Yablonovitch1994}, especially in two-dimensional (2D) periodic dielectric
structures, also known as photonic crystal slabs \cite{Reese2001}. These
structures offer the possibility of guiding light along their extended
(non-periodic) dimension and are much more easily constructed than full
three-dimensional (3D) photonic crystals.

Micro-machined photonic bandgap crystals have been both fabricated and
experimentally characterized \cite{Ozbay1996}. Several authors have
performed theoretical calculations of bandgaps in these materials \cite%
{Birks1995,Joannopoulos1999}, as well as experimental studies of the
transmission through a 2D photonic crystal slab at a wavelength of 1.55 nm %
\cite{Joannopoulos2000}. In addition, applications of these and other
similar kinds of photonic crystals to microwave mirrors, substrates for
planar antennas, and photonic crystal heterostructures, have already been
proposed and studied \cite{Malloy1996}. A one-dimensional photonic crystal
bandgap has also recently been observed in the radio-frequency region \cite%
{Hache2002}.

In the visible and near-infrared part of the spectrum, there has been much
recent progress in the fabrication of 2D ``photonic crystal fibers'' \cite%
{Cregan1999}. One type of these optical fibers consists of 2D
hexagonal-close-packed glass structures, fabricated by stretching out a
macroscopic hexagonal lattice of heated, hollow glass cylinders into a
microscopic glass structure of the same geometry. These structures can also
be constructed with a central ``defect,'' such as a hexagonal central hole
consisting of seven nearest-neighbor hollow cylinders removed from the
center of the structure. The resulting hexagonally shaped central hole can
serve as a waveguide for electromagnetic radiation, since the surrounding
periodic dielectric structure (cladding) possesses a photonic bandgap.
Inside the cladding structure only evanescent wave solutions should be able
to propagate transversely to the central axis of the structure.

As a first step towards the study of the properties of these 2D optical
waveguide structures, we have performed some measurements of an analogous
dielectric structure in the microwave region of the electromagnetic
spectrum. Since Maxwell's equations are scalable, it should be possible to
directly apply the results obtained in the microwave region to structures
designed for different wavelengths, such as the visible and near infrared.
There are at least two significant reasons for conducting these studies with
microwave-scale structures. First, their large scale allows relatively easy
fabrication and characterization. Second, without any scaling, these
low-loss periodic dielectric structures could be relevant for electron beam
devices at microwave and millimeter wavelengths (e.g., travelling-wave tubes
and backward-wave oscillators). This is especially significant since
electron beams could interact strongly with electromagnetic waves at these
wavelengths when they co-propagate either longitudinally down a central
hole, or transversely across the top surface of these periodic dielectric
structures\cite{NLGW2001}. One possible future application of hollow-core
photonic crystal fibers is the acceleration of relativistic electrons to
high energies using short-duration laser pulses co-propagating along with a
relativistic electron beam inside the fiber. The results of our measurements
should be directly relevant to the design of these kinds of electron beam
devices\cite{capri2000}.

Theoretical calculations of the bandgap structure of periodic holes in a
hexagonal array have been carried out by Johnson et al\cite%
{Joannopoulos1999,Foteinopoulou2001}. The structure used in our experiment
is similar to the one considered in these calculations, except for the
presence of small, triangular, interstitial holes between each trio of
adjacent acrylic rods, which we believe are negligible for the determination
of the bandgap. The Johnson calculations show that for a lattice constant $a$%
, hole diameter $d=0.9a$, and slab thickness $0.6a$, the bandgap should be
located at $\nu _{\mathrm{bg}}\approx 0.4c/a$. For our system this implies $%
\nu _{\mathrm{bg}}\approx 10$\ GHz. Based on optical fiber experiments \cite%
{Cregan1999}, we expect that a photonic bandgap should be present in our
structure around 12 GHz.

\begin{figure}[tbp]
\centerline{\includegraphics{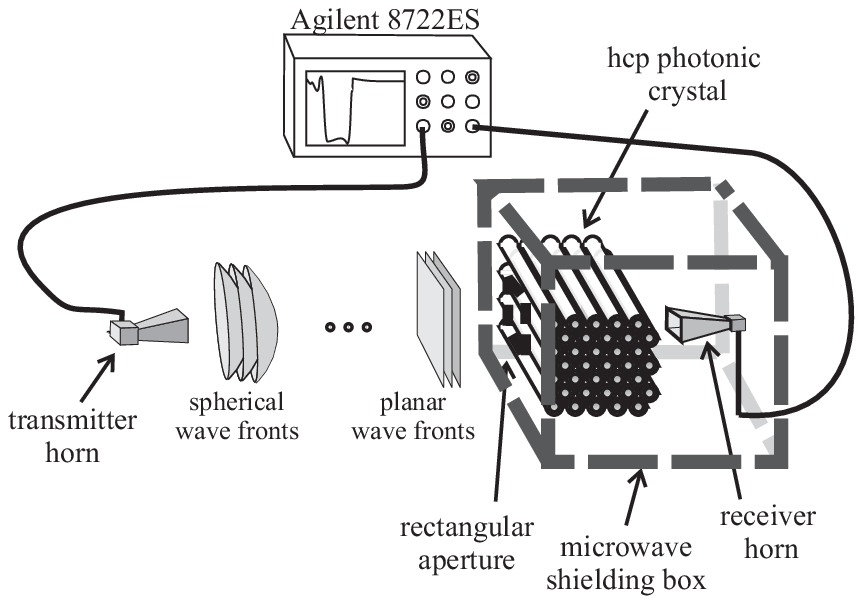}}
\caption{Experimental setup for measuring the photonic bandgap of the
photonic crystal slab.}
\label{setup}
\end{figure}

In Fig. \ref{setup} we show a schematic of the experimental apparatus used
for the microwave transmission measurements. The photonic crystal slab was
constructed by stacking together acrylic pipes (23\thinspace cm long,
1/2\thinspace \thinspace inch outer diameter, 3/8\thinspace \thinspace inch
inner diameter) in a hexagonal lattice, resulting in a structure with an
air-filling fraction of approximately 0.60. \ The pipes were then glued
together at their surfaces of contact using a standard acrylic cementing
solvent. \ The dielectric constant of acrylic (polymethyl-methacrylate) at
10 GHz is 2.59\cite{Gray1972}. The microwaves were generated by a commercial
microwave network analyzer (Agilent 8722ES) connected to a microwave
transmitter horn. The analyzer was swept from 8 to 14\thinspace\ GHz in 15
MHz increments. The slab and receiver horn were placed 1.6 \thinspace m away
in a box measuring 60\thinspace\ cm on a side, with microwave-absorbing
walls. Microwaves entered the box through a 14 cm $\times $ 17 cm
rectangular aperture. These dimmensions were chosen to minimize diffraction
effects through the aperture without allowing leakage around the slab. We
normalized transmission measurements by removing the slab and measuring the
total microwave signal at the receiving horn. With the slab removed and the
rectangular aperture closed, the microwave signal at the receiver is
suppressed by more than 45 dB, indicating good shielding by the box.

\begin{figure}[tbp]
\centerline{\includegraphics{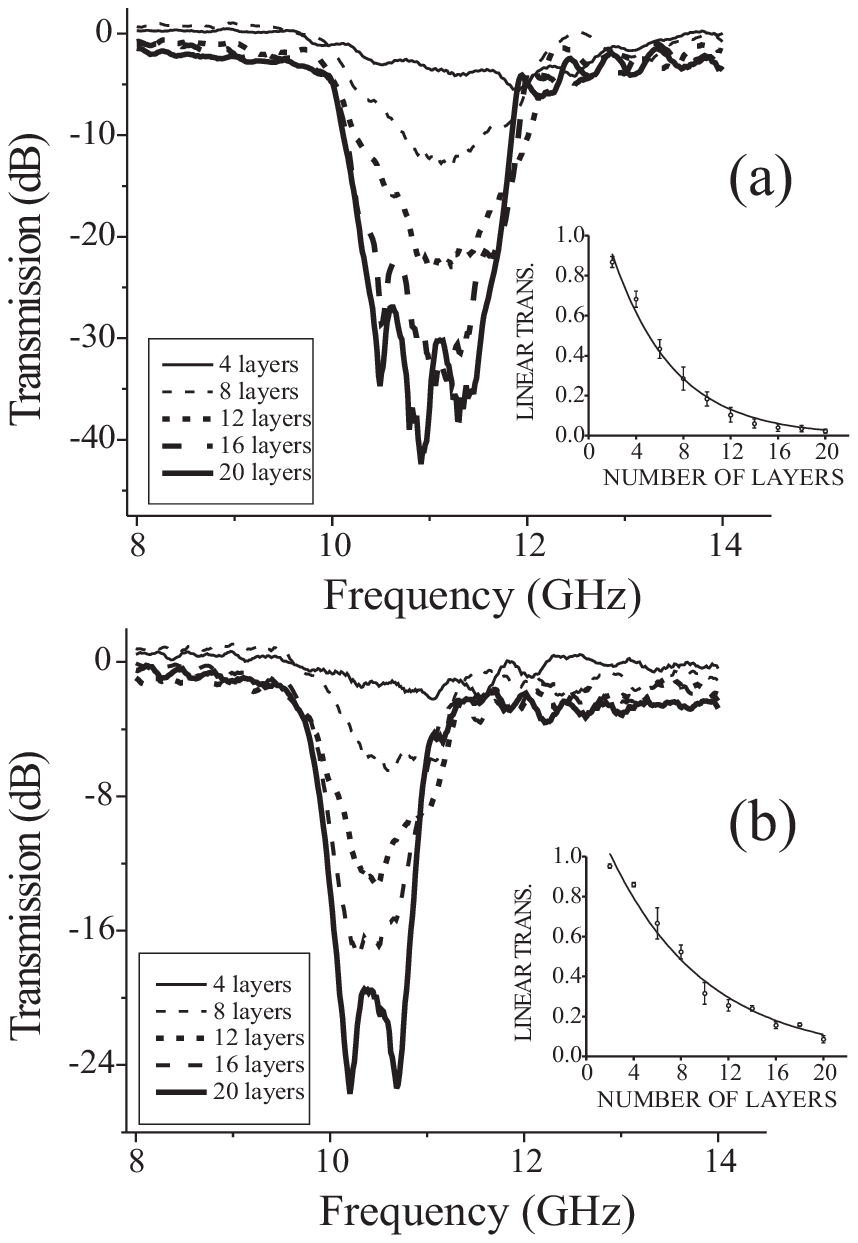}}
\caption{Microwave transmission versus frequency for electric field (a)
perpendicular (TM) and (b) parallel (TE) to the dielectric rods, for various
numbers of slab layers. The inset shows the average transmission versus
number of slab layers in a frequency window from (a) 10.5 GHz to 11.5 GHz
and (b) 10.25 GHz to 10.75 GHz.}
\label{layertrans}
\end{figure}

In Fig. \ref{layertrans} we show the normalized microwave transmission as a
function of frequency for various numbers of photonic crystal slab layers.
For the case in which the electric field is perpendicular to the dielectric
rods [transverse magnetic or (TM)], there is a clear bandgap centered at 11
GHz whose depth increases with the number of layers. The bandgap is about
1.5 GHz wide and its width and center frequency appears to be independent of
layer number. Note that the minimum transmission value is above our
background transmission, indicating that these measurements are above the
noise. For the case of the electric field parallel to the rods [transverse
electric or (TE)], we find a narrower ($\approx $ 1 GHz), shallower bandgap
centered at 10.5 GHz. To quantify the bandgap depth, we average the
transmission value for each slab layer number, in a frequency window from
10.5 GHz to 11.5 GHz in the TM case and 10.25 GHz to 10.75 GHz in the TE\
case. Fitting the results to an exponential decay, we find a $1/e$ power
decay length of 5.2 layers and 8.1 layers, respectively.

\begin{figure}[tbp]
\centerline{\includegraphics{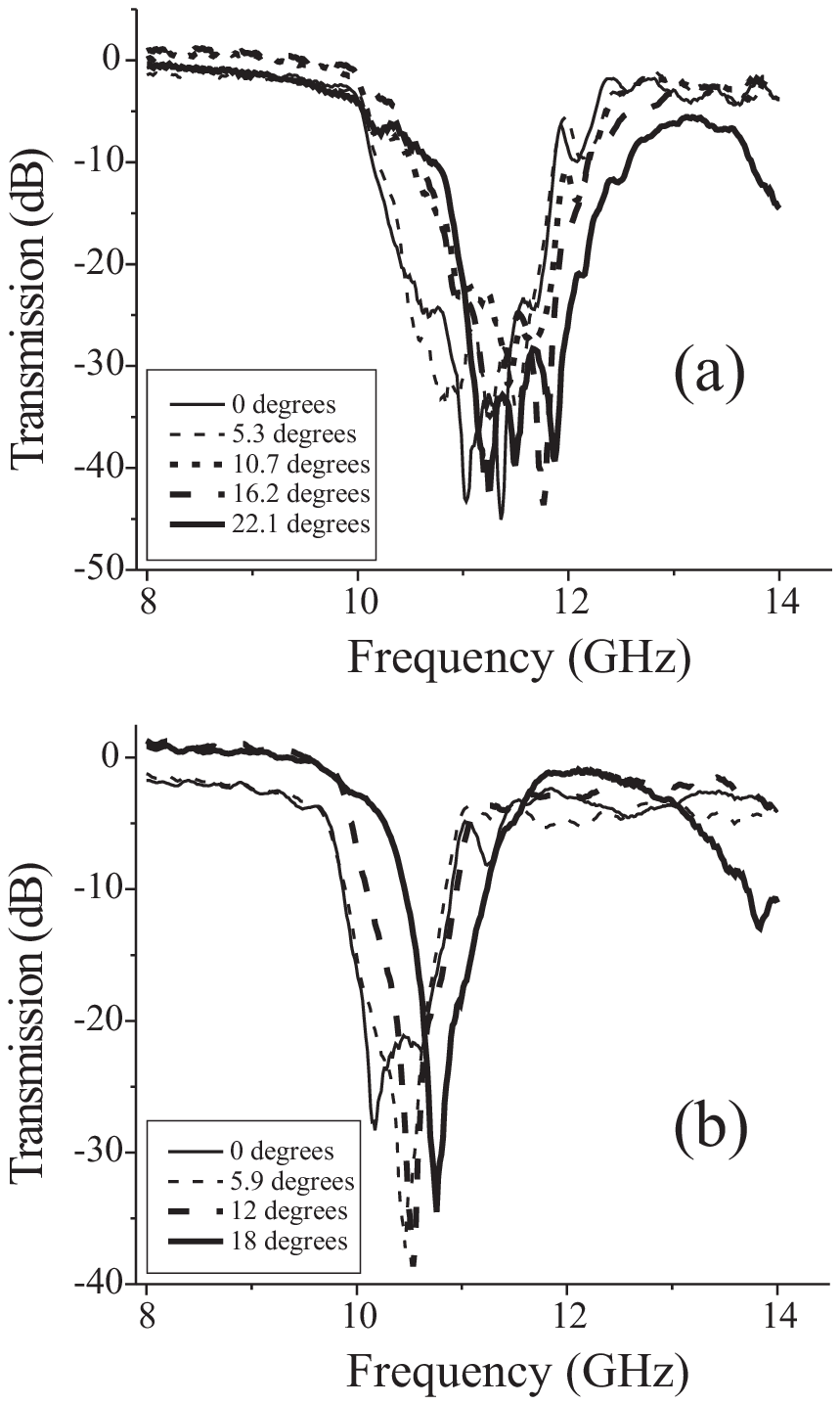}}
\caption{Microwave transmission versus frequency for several angles. (a)
Electric field perpendicular to rods (TM). (b) Electric field parallel to
rods (TE).}
\label{transangle}
\end{figure}

We have also taken transmission measurements at various angles of incidence,
using a 20-layered slab. The results are shown in Fig. \ref{transangle} for
a plane of incidence perpendicular to the dielectric rods (i.e. no
wave-vector component perpendicular to the plane of periodicity). As was
previously seen in Fig. \ref{layertrans}, the configuration TM bandgap is
deeper and wider than the TE bandgap. The angle data show that for both
polarizations a larger angle of incidence shifts the bandgap to higher
frequencies. Using the -15 dB points as a reference, a higher angle of
incidence produces a narrower bandgap in the TE case but not in the TM case.

In conclusion, we have demonstrated the existence of a photonic bandgap in a
2D hexagonal photonic crystal slab. Its depth depends on the number of
layers, with an exponential decay of transmission on the order of several
layers. We have also measured the dependence of transmission on the angle of
incidence and polarization of the incoming radiation, finding that the
bandgap persists at high angles but is shifted and in certain cases narrowed.

This work was supported by ARO.

\pagebreak

\pagebreak 

\end{document}